\documentclass[preprint,12pt,authoryear]{elsarticle}
\usepackage{amsmath}
\usepackage{graphicx}
\usepackage{booktabs}
\usepackage[table]{xcolor}
\usepackage{listings}
\usepackage{enumerate}
\usepackage{natbib}
\usepackage[a4paper]{geometry}
\usepackage{graphicx}
\usepackage[textwidth=8em,textsize=small]{todonotes}
\usepackage{amsmath}
\usepackage{natbib}
\usepackage{array}
\usepackage[utf8]{inputenc}
\usepackage{graphicx}
\usepackage{longtable}
\graphicspath{ {./images/} }
\usepackage{multirow}
\usepackage{float}
 \usepackage{xspace}
 \usepackage{appendix}
 \usepackage{array}
\usepackage{amssymb}
\usepackage{dirtytalk}
\usepackage{etoolbox}
\usepackage{url}
\newcolumntype{C}[1]{>{\centering\arraybackslash}p{#1}}

\usepackage{booktabs}

\usepackage{url} 
\date{}
\journal{ARXIV}

\begin{document}

\begin{frontmatter}

\title{Forecasting Mortality Rates: Unveiling Patterns with a PCA-GEE Approach}
\author{Reza Dastranj\corref{cor1}}

\ead{dastranj@math.muni.cz}

\author{Martin Kol\'a\v r}
\ead{mkolar@math.muni.cz}

\address{Department of Mathematics and Statistics, Masaryk University, Kotlářská 2, 611 37 Brno, Czech Republic}

\cortext[cor1]{Corresponding Author}

\begin{abstract}

Principal Component Analysis (PCA) is a widely used technique in exploratory data analysis, visualization, and data preprocessing, leveraging the concept of variance to identify key dimensions in datasets. In this study, we focus on the first principal component, which represents the direction maximizing the variance of projected data. We extend the application of PCA by treating its first principal component as a covariate and integrating it with Generalized Estimating Equations (GEE) for analyzing age-specific death rates (ASDRs) in longitudinal datasets. GEE models are chosen for their robustness in handling correlated data, particularly suited for situations where traditional models assume independence among observations, which may not hold true in longitudinal data. We propose distinct GEE models tailored for single and multipopulation ASDRs, accommodating various correlation structures such as independence, AR(1), and exchangeable, thus offering a comprehensive evaluation of model efficiency. Our study critically evaluates the strengths and limitations of GEE models in mortality forecasting, providing empirical evidence through detailed model specifications and practical illustrations. We compare the forecast accuracy of our PCA-GEE approach with the Li-Lee and Lee-Carter models, demonstrating its superior predictive performance. Our findings contribute to an enhanced understanding of the nuanced capabilities of GEE models in mortality rate prediction, highlighting the potential of integrating PCA with GEE for improved forecasting accuracy and reliability.

\end{abstract}


\end{frontmatter}

\section{Introduction}\label{sec:intro}

Actuarial mortality tables have been a cornerstone in mortality studies. In the 1800s, Benjamin Gompertz observed a pattern in these tables, suggesting that the probability of death increases exponentially with age \citep{gompertz1825xxiv}. This paved the way for further advancements. The LC model \citep{lee1992modeling}, developed in the 1990s, stands out as a significant contribution. The LC model's simplicity and effectiveness made it a popular foundation for numerous extensions in mortality forecasting.

While the LC model excels in single-population forecasting, recent developments are pushing the boundaries towards multi-population scenarios \citep{li2005coherent}. The rationale behind this is that incorporating data from multiple populations can enhance the reliability of parameter estimation. This, in turn, leads to more accurate forecasts. Researchers have explored various approaches to achieve this, including introducing common effects across populations and utilizing neural networks \citep{kleinow2015common, richman2021neural}. These advancements hold promise for improving the accuracy of mortality projections in an increasingly interconnected world. For a review of the mortality modelling and forecasting methods, see \cite{booth2008mortality}.

The 20th century witnessed a significant decline in global mortality rates, leading to a steady increase in life expectancy. There is no clear indication that this trend of increasing life expectancy has reached its maximum \citep{burger2012human}. However, this prolonged increase in life expectancy has placed additional strain on support systems catering to the elderly, including government services, healthcare systems, life insurance providers, and pension funds. 

In particular, the pricing and reserving of annuities within the pension funds industry and insurance sector rely on projections of future ASDRs, which signify the mortality risk at different ages. This heightened emphasis on longevity underscores the need for the development of robust models capable of accurately projecting mortality rates.

ASDRs serve as vital indicators of mortality trends, collected sequentially over different years for various populations. Within the same age group, mortality rates exhibit higher similarity, forming a longitudinal dataset \citep{frees2004longitudinal} with inherent correlations \citep{dastranj2023age}.

Traditional regression models like linear or nonlinear regression may not be suitable for analyzing longitudinal data, particularly when dealing with age groups as subjects. This is because such data often include repeated measurements within age groups and potential correlations over time. To address these complexities and ensure accurate estimation of variance and p-values, specialized statistical approaches like GEE models are required.

In our study, we aim to develop a GEE model for forecasting mortality rates, leveraging a covariate denoted as $k_{ct}$. This covariate plays a crucial role in capturing a significant percentage of variation in mortality rates within our study context. Specifically, $k_{ct}$ captures a substantial portion of the variation in mortality rates across three distinct age groups: 0 to 19, 20 to 50, and 51 to $\omega$, for both females and males within each country. To categorize individuals into these age groups, we consider them homogeneous within each group. These age categories are delineated as children (0 to 19), young adults (20 to 50), and older adults (51 to $\omega$), based on an analysis of key factors influencing mortality rates across different age demographics.

To further refine our analysis, we conduct PCA \citep{jolliffe2002principal,abdi2010principal} separately for children, young adults, and older adults within each country in our dataset. The covariate $k_{ct}$ is derived as the first principal component (PC1) obtained from PCA applied to the logarithm of mortality rates across these age groups. Utilizing the GEE model, we can effectively predict and understand the variation in the logarithm of the ASDRs variable, denoted as $y_{cgxt}$, based on the covariate $k_{ct}$. The proportion of variation accounted for by the first principal component indicates the extent to which PC1 captures the variability in mortality data. Our analysis reveals that PC1 explains approximately 99 percent of the variability in ASDRs. By integrating GEE modelling with PCA-derived covariates, our approach offers a robust framework for forecasting mortality rates while accounting for the complex dependencies present in longitudinal data across different age groups.

Traditional statistical approaches, such as generalized linear models (GLMs), assume independence among individual rows in the data \citep{nelder1972generalized, mccullagh2019generalized,james2013introduction}. However, when dealing with longitudinal and clustered data, especially in mortality rates within the same age group, this assumption falls short, violating the independence and identically distributed requirement. To address these limitations, statisticians developed methods, leading to the emergence of GEE as a powerful tool explicitly designed to extend the GLM algorithm. GEE provides a robust framework for modelling correlated data, particularly in scenarios where straightforward GLM methods may fall short \citep{liang1986longitudinal, hardin2002generalized,lee2004conditional}.

This paper delves into the powers and properties of GEE in the context of mortality forecasting. Two distinct models are presented: one tailored for multipopulation scenarios and another designed for single-population studies. The multipopulation model incorporates key predictors such as country, gender, age, and their intricate interactions with mortality covariates $k_{ct}$, $k_{ct}^2$, and cohort.

In the landscape of statistical modelling, linear mixed effects models (LMEs) and GEEs represent two prominent approaches for handling correlated data.

Subject-Specific Models (LMEs) focus on estimating both fixed effects and variance components, providing insights into the magnitude of each source of variation in the data \citep{pinheiro2006mixed}. They are particularly suitable when there is interest in understanding the underlying population and capturing the variability attributed to different grouping factors. However, subject-specific models have a conditional formulation with random intercepts and slopes, making it challenging to provide a population-average interpretation of the model parameters. Estimates obtained from subject-specific models are specific to the individuals or subjects under consideration \citep{lee2004conditional}.

\cite{dastranj2023age} have introduced an LME model for modeling and forecasting mortality rates. In this model, the inclusion of ``country'' as a random effect \citep{gelman2005analysis, green1960complex} indicates a specific interest in the underlying populations of the six European countries included in the dataset. The variance of the random effect for the country quantifies the variability in the response variable attributed to differences between these observed countries. When exploring the variance of the random effect for countries beyond the dataset, the analysis considers how much the response variable may vary for countries not explicitly observed. This approach facilitates a broader understanding of the potential variability in mortality rates across a larger population, extending insights beyond the specific countries included in the dataset.

Marginal Models (GEEs), on the other hands, prioritize estimating regression coefficients without explicit interest in variance components. They are robust models designed to handle correlated data and are especially useful when the primary concern is modelling the mean response while treating within-group correlation as a nuisance. Marginal models aim to capture the population-average effects \citep{lee2004conditional}, allowing for a more generalized interpretation of the model parameters. However, they may encounter challenges when dealing with ordinal data, particularly in specifying an underlying distribution. This is especially evident when GEEs are applied, as they focus on estimating population-average effects while accommodating correlated data.

GEEs require specifying a correlation structure for repeated measures. Several commonly used structures include:

\paragraph{Independence}
The Independence structure assumes no correlation between repeated measures, similar to ordinary least squares. The corresponding correlation matrix is represented as:
\[
\begin{bmatrix}
1 & 0 & \cdots &0 \\
0 & 1 & \cdots & 0 \\
\vdots &\vdots &\ddots &\vdots\\
0 & 0 &  \cdots &1 \\
\end{bmatrix}
\]

The method of independence estimating equations incorrectly assumes that observations within a subject are independent, treating correlated responses as if they were independent \citep{pan2002selecting}. When specifying an independence correlation structure for repeated measures in GEEs, it essentially assumes zero correlation between repeated measures. This implies independence, rendering the GEE model behavior akin to a GLM. The specification of an independence correlation structure treats each observation as entirely independent of others, resembling a standard GLM. It neglects potential correlations or dependencies between observations, resulting in a simpler and more straightforward model concerning the correlation structure.

\paragraph{AR(1)}
The Autoregressive (AR(1)) structure assumes decreasing correlation with increasing time separation. The corresponding correlation matrix is:
\[
\begin{bmatrix}
1 & \rho & \cdots&\rho^{n-1} \\
\rho & 1 & \cdots &\rho^{n-2} \\
\vdots &\vdots &\ddots &\vdots\\
\rho^{n-1} & \rho^{n-2} & \cdots&1 \\
\end{bmatrix}
\]

\paragraph{Exchangeable}
Exchangeable, or compound symmetry, assumes all correlations between repeated measures are the same. The correlation matrix for this structure is given by:
\[
\begin{bmatrix}
1 & \rho & \cdots&\rho \\
\rho & 1 & \cdots&\rho \\
\vdots &\vdots &\ddots &\vdots\\
\rho & \rho & \cdots&1 \\
\end{bmatrix}
\]

\paragraph{Unstructured}
The Unstructured structure is the most flexible, allowing different correlations between all pairs of repeated measures. The correlation matrix is represented as:
\[
\begin{bmatrix}
1 & \rho_{12} & \cdots& \rho_{1n} \\
\rho_{21} & 1&\cdots & \rho_{2n} \\
\vdots &\vdots &\ddots &\vdots\\
\rho_{n1} & \rho_{n2} &\cdots& 1 \\
\end{bmatrix}
\]

Each $\rho$ represents a correlation coefficient, capturing the strength and direction of the correlation between respective measures. These structures provide flexibility in modelling the dependence between repeated measures, catering to different assumptions about the nature of the correlation in longitudinal data.

In comparing the fit of GEE models with different correlation structures, the Quasi-Likelihood Information Criterion (QIC) serves as a robust measure. QIC balances goodness of fit and model complexity, guiding the selection of the most suitable model. Unlike traditional criteria like Akaike Information Criterion (AIC) and Bayesian Information Criterion (BIC), which may not align well with GEE's quasi-likelihood framework, QIC is specifically tailored for accurate model comparison in this context. Its variants, like QICu, offer additional corrections for overdispersion. Therefore, when dealing with GEE models for mortality data, QIC stands out as a more appropriate and reliable criterion, ensuring precise model selection.

The foundational model structure for GEEs in the context of mortality rates and age groups can be expressed as:

\begin{equation}
  y_{xt} = \beta_0 + \sum_{k} X_{xtk}\beta_k + \text{CORR} + \text{error}.  
\end{equation}

Within this formulation, $\beta$'s represent fixed effects, while ``CORR'' encapsulates the correlation structure inherent in GEE. For mortality rates within specific age groups, we assume we have $y_{xt}$ (log-transformed mortality rates) as our outcome for individual age group $x$ at time $t$. This could just as easily represent individuals within other types of levels, like genders or counties. Additionally, we have $X_{xt}$, the matrix of predictors. 

In the context of repeated observations within age groups over time, it is crucial to address the potential biases in standard error estimations for both time-varying and time-invariant predictors. Specifically, conventional models not incorporating the correlation within age groups over time tend to exhibit an overestimation of standard errors for $\beta$'s associated with time-varying predictors. Simultaneously, there is a tendency for the standard errors of time-invariant predictors to be underestimated. This phenomenon underscores the necessity of employing advanced modelling techniques, such as GEE, which explicitly account for temporal dependencies and yield more accurate standard error estimates. GEE's capacity to appropriately handle the correlation structure within longitudinal data, especially within age groups, enhances the reliability of parameter estimates, contributing to robust statistical inference in the presence of repeated observations.

\section{Modelling Mortality Rates in Multi-Population}\label{sec2}

We propose two approaches to define a covariate $k_{ct}$ for modelling and forecasting mortality rates. The first approach, referred to as PCA-GEE (Principal Component Analysis - Generalized Estimating Equations), involves using principal component analysis to derive the covariate $k_{ct}$. The second approach, termed Avg-GEE (Average - Generalized Estimating Equations), involves averaging mortality rates across predefined age groups to derive the covariate $k_{ct}$ (see \cite{dastranj2023age}). In our study, we will consider both PCA-GEE and Avg-GEE approaches and compare their forecast accuracy with that of the LL model. By evaluating the performance of these approaches, we aim to identify the most effective method for modelling and forecasting mortality rates.

We now begin to elaborate on the establishment of the PCA-GEE method. Let $q_{cgx}=1-exp(-d_{cgx}/e_{cgx})$ denote the ASDR at age $x$ and time $t$ of gender $g$ in country $c$, for $c=$ $1, 2,\cdots,M$; $g=1, 2$; $x = 0, 1, \cdots , \omega$; and $t=0,1, \cdots, T$. Let $y_{cgxt}=\log (q_{c,g,x,t})$. 

We categorize individuals into three distinct age groups: 0 to 19, 20 to 50, and 51 to $\omega$, treating each group as homogeneous. These age categories are identified as children, young adults, and older adults, respectively. This classification is based on an analysis of significant factors influencing mortality rates across different age demographics.

We conduct PCA on the children, young adults, and older adults of each country within the dataset. Subsequently, we define covariate $k_{ct}$ as follows:

\begin{align}
k_{ct} = &
\begin{cases} 
{\rm PC1}_{ct}^{[0,19]} & \text{if}\ x  \in [0,19], \\ 
{\rm PC1}_{ct}^{[20,50]} & \text{if}\ x  \in [20,50], \\ 
      {\rm PC1}_{ct}^{[51,\omega]} & {\rm otherwise},
\end{cases}
\label{ali5}
\end{align}
for $c=1, 2, \cdots, M$, and $t=0, 1,  \cdots, T$.

$k_{ct}$ will serve as a predictor in our PCA-GEE model. To forecast future values of $k_{ct}$, we will employ random walks with drift \citep{box2015time} for ages ranging from 0 to 19 within each country, as well as for ages 20 to 50 and ages 51 to $\omega$, also within each country. This covariate, termed the ``mortality covariate,'' holds significance in capturing the temporal patterns of mortality rates. We propose the development of a GEE model to capture the relationships between the log ASDRs and the mortality covariate. Within this framework, the mortality covariate is regarded as the driver or exogenous series, while the log ASDRs are considered as dependent variables influenced by this driver.

The PCA-GEE model for multi-population scenarios, implemented through the geeglm function available in the geepack package for R \citep{hojsgaard2006r}, is defined as follows:

\begin{align}
    \text{geeglm}(y \sim & \text{country}+\text{gender}+\text{age} + \nonumber \\ 
       & \text{gender:age:I}(k_{ct}) + \text{gender:age:I}(k_{ct}^2) + \text{cohort}, \nonumber \\
    &  \text{id} = \text{country:gender:age}, \text{waves} = \text{year}, \nonumber \\
    &  \text{corstr} = \text{``ar1''}, \text{weights} = \text{agenum}/\text{mean(agenum)}, \nonumber \\
    &  \text{data} = \text{ASDRs}) \nonumber
\end{align}

This model incorporates various predictors, including age, country, gender, and their interactions, to capture the complexity of mortality rates within the specified populations. The selection of the correlation structure (corstr) is pivotal in accounting for temporal dependencies within the data, specifically recognizing the potential correlation of mortality rates over consecutive years. By choosing an appropriate correlation structure, such as autoregressive (``ar1''), the models acknowledge and adjust for these temporal dependencies. This consideration significantly enhances the reliability of parameter estimates, contributing to robust and accurate mortality forecasting in both single and multi-population scenarios. The corstr argument in the geeglm function is a character string that designates the correlation structure for the GEE model. The permissible options include: ``independenc'',
``exchangeabl'',
``ar1'',``unstructured'', and ``userdefined'' (allows the specification of a user-defined correlation structure).

The ``id'' parameter defines the grouping structure within the data, serving as a unique identifier for each observation. For instance, specifying id = country:gender:age implies that observations are uniquely identified by a combination of country, gender, and age. The ``waves'' parameter signifies the time points or waves of data collection, highlighting the longitudinal aspect of the data. The ``weights'' parameter represents the applied weighting scheme for observations. In this context, the term agenum refers to treating age as a numeric variable. However, within the model, the term age is considered as a factor.

\cite{dastranj2023age} have demonstrated the rationale behind incorporating the quadratic term ($k_{ct}^2$) in the model when utilizing the mortality covariate $k_{ct}$. It revealed that the relationship between $y_{xt}$ and $k_{ct}$ exhibits a quadratic trend, suggesting a more accurate representation of the underlying dynamics. The plotted graph of $y_{xt}$ against $k_{ct}$ revealed a combination of a quadratic trend and stochastic variation, reinforcing the justification for introducing the quadratic term in the model.

The PCA-GEE model for the multi-population scenario can be mathematically represented as:
\begin{equation}
    y_{cgxt} =  a_c+a_g+a_x + b_{gx}k_{ct}+ c_{gx}k_{ct}^2 + \gamma(t-x) + \epsilon_{cgxt}
\end{equation}
In this formulation, the terms $a_c$, $a_g$, and $a_x$ represent the country, gender, and age effects in the intercept, respectively. Furthermore, $b_{gx}$ and $c_{gx}$ indicate the gender-age effects of $k_{ct}$ and $k_{ct}^2$, while $\gamma$ denotes the cohort effect. The error term is denoted by $\epsilon_{cgxt}$.

We now commence the establishment of the Avg-GEE method. Let $k_{ct}$ denote the average of $y_{cgxt}$  across three distinct age range groups, encompassing both females and males in country $c$ at time $t$. These groups include the average for age groups 0 to 19, the average for age groups 20 to 50, and the average for age groups 51 to $\omega$:

 \begin{align}
k_{ct} &= 
\begin{cases} 
k_{ct}^{[0,19]} & \text{if } x \in [0,19], \\ 
k_{ct}^{[20,50]} & \text{if } x \in [20,50], \\ 
k_{ct}^{[51,\omega]} & \text{otherwise},
\end{cases}
\label{eq:avg_kct}
\end{align}

where \( c = 1, 2, \ldots, M \) and \( t = 0, 1, \ldots, T \). 

The average values \( k_{ct}^{[0,19]} \), \( k_{ct}^{[20,50]} \), and \( k_{ct}^{[51,\omega]} \) are calculated as follows:

\begin{align*}
k_{ct}^{[0,19]} &= \frac{1}{2 \cdot 20} \sum_{g=1}^{2} \sum_{x=0}^{19} y_{cgxt}, \\
k_{ct}^{[20,50]} &= \frac{1}{2 \cdot 31} \sum_{g=1}^{2} \sum_{x=20}^{50} y_{cgxt}, \\
k_{ct}^{[51,\omega]} &= \frac{1}{2 \cdot (\omega-50)} \sum_{g=1}^{2} \sum_{x=51}^{\omega} y_{cgxt},
\end{align*}
where \( c = 1, 2, \ldots, M \) and \( t = 0, 1, \ldots, T \), and \( \omega \) represents the upper limit of the age range for country \( c \) at time \( t \). 
 
 $k_{ct}$ will be included as predictor in our Avg-GEE model. We will utilize random walks with drift to predict the future values of $k_{ct}$ for ages $0$ to $19$ within each country, as well as the future values of $k_{ct}$ for ages $20$ to $50$ and ages $51$ to $\omega$, also for each country. 

The Avg-GEE model for multi-population scenarios, implemented through the geeglm function available in the geepack package for R (Højsgaard et al., 2006), is defined similarly to the PCA-GEE model. However, in the Avg-GEE model, we utilize the covariate $k_{ct}$, as defined in Equation \ref{eq:avg_kct}, obtained by taking the average of log mortality rates across predefined age groups.

In Figure \ref{fig1}, the thick blue curve corresponding to the Czech Republic depicts $k^{[51,80]}_{Czech \ t}$, which is derived by averaging the logarithm of mortality rates across ages 51 to 80 for both males and females. For instance, $k^{[51,80]}_{Czech \ 1991}$ represents the average log ASDRs of both genders in the Czech Republic in 1991. $k^{[51,80]}_{Czech \ t}$ denotes the aggregated age-specific mortality pattern across ages 51 to 80 for both males and females in the Czech dataset \citep{HMD}.
\begin{figure}[H]
    \centering
    \includegraphics[width=\textwidth]{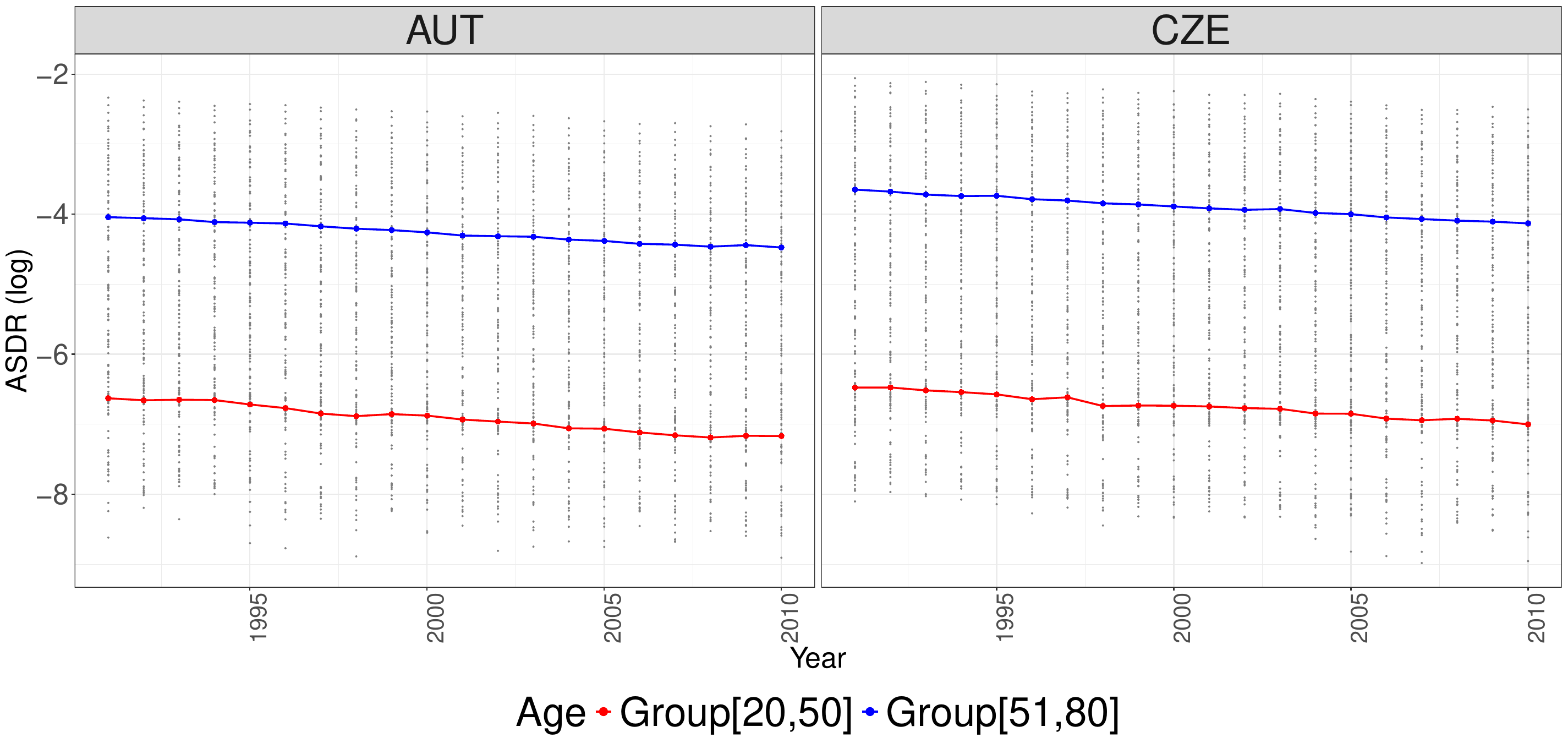}
    \caption{This plot depicts the average of logarithmized ASDRs across age groups 20 to 50 and 51 to 80 for both females and males in Austria (AUT), as well as the average of logarithmized ASDRs across the same age groups for both genders in the Czech Republic (CZE). The data spans the years 1991 to 2010.}
    \label{fig1}
\end{figure}

\subsection{Multi-Population Analysis - Austria and Czech Republic (Females and Males, Age 20-80)}

This section explores mortality rates for Austria and Czech Republic females and males aged 20 to 80 during 1991-2010, serving as the training set.

The proportion of variance explained by both $\rm{PC1}^{[20,50]}$ and $\rm{PC1}^{[51,80]}$ for both  Austria and Czech Republic amounts to 99 percent. This indicates that $\rm{PC1}^{[20,50]}$ and $\rm{PC1}^{[51,80]}$ each capture a very high amount of the total variability in the mortality rates for their respective age groups. They are both the first principal components for their respective age ranges and contain a large amount of information about the mortality rates across these age ranges, summarizing most of the important patterns and trends in the data. By retaining both $\rm{PC1}^{[20,50]}$ and $\rm{PC1}^{[51,80]}$, we effectively reduce the dimensionality of the data while still retaining the majority of the relevant information, simplifying the modelling process and making it more computationally efficient. These components are linear combinations of the original variables and can be interpreted as representative components that capture the overall trend or pattern in the mortality rates for their respective age groups.

\begin{figure}[H]
    \centering
    \includegraphics[width=\textwidth]{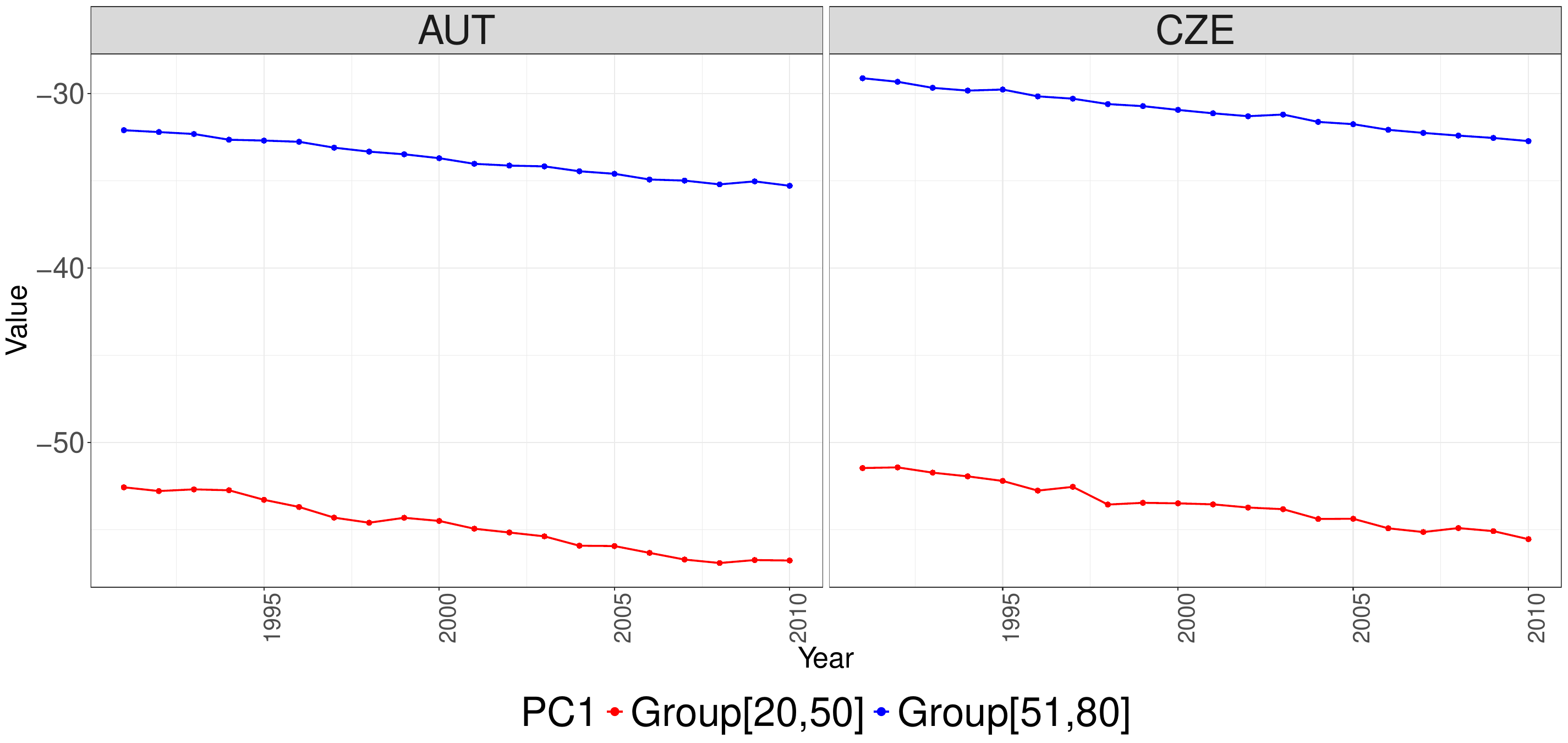}
    \caption{This plot illustrates the PC1 obtained from PCA performed separately for age groups 20 to 50 and 51 to 80 for both females and males in Austria (AUT), as well as for both genders in the Czech Republic (CZE). Therefore, there are four curves on the plot, each representing the PC1 obtained from PCA for the specified age groups and countries. The data covers the years 1991 to 2010.}
    \label{fig1}
\end{figure}

Utilizing three GEE models (geeInd, geeEx, geeAr1) with independence, exchangeable, and AR(1) correlation structures, Table \ref{tab2} displays QIC values for model selection. 

\begin{table}[H]
  \centering
  \caption{QIC Values for Models with Different Correlation Structures (Multipopulation).}
  \label{tab:qic-values}
  \begin{tabular}{lcccccc}
    \toprule
    \textbf{Model} & \textbf{QIC} & \textbf{QICu} & \textbf{Quasi Lik} & \textbf{CIC} & \textbf{Params} & \textbf{QICC} \\
    \midrule
    geeInd (PC1) &  532.3  &    677.5  &    -30.8 &     235.4  &    308 &   -2396.1  \\
    geeEx (PC1) & 624.4  &   677.9  &   -30.9  &   281.2 &    308  & -2278.3  \\
    geeAr1 (PC1) & 538.8  &   677.5  &   -30.8 &    238.6 &    308 &  -2363.9  \\
    geeInd (Avg)&  744.6  &   677.3  &   -30.6 &    341.6 &    308 &  -2183.8   \\
    geeEx (Avg)& 894.7   &  677.7 &    -30.8    & 416.5 &    308&   -2008.0 \\
    geeAr1 (Avg)& 756.2  &   677.3   &  -30.6 &    347.5   &  308&   -2146.5 \\
    \bottomrule
  \end{tabular}
   \label{tab2}
\end{table}

Opting for geeAr1, incorporating an AR(1) structure, this final model predicts mortality rates for the four populations during 2011-2019.

To evaluate predictive accuracy, we consider the mean square error (MSE). Table \ref{tab3} displays the MSE values for the test set (years $2011-2019$) across the four populations. We employed the MultiMoMo package (Copyright (c) 2020 Katrien Antonio; Sander Devriendt; Jens Robben) to fit the LL model to the dataset of the four populations. Subsequently, we employ the demographic LL model as a benchmark to assess the performance of the GEE model. 

In the test datasets, the MSEs of the PCA-GEE model are consistently lower than those of the LL model across all four populations. Similarly, the MSEs of the Avg-GEE model also exhibit lower values compared to the LL model across the same four populations. These findings indicate that, on average, both the PCA-GEE and Avg-GEE models outperform the LL model in predicting mortality rates across these populations.

\begin{table}[H]
\small
\centering
\begin{tabular}{|c|c|c|c|c|}
\hline
& \multicolumn{2}{c|}{AUT} & \multicolumn{2}{c|}{CZE} \\
Error & Female & Male & Female & Male \\
\hline
LL test set  & 1.37$\times10^{-6}$ & 1.87$\times10^{-6}$ & 6.33$\times10^{-7}$ & 3.65$\times10^{-6}$  \\
GEE (PC1) test set &  \cellcolor{green!25}1.09$\times10^{-6}$& \cellcolor{green!25} 1.82$\times10^{-6}$ & \cellcolor{green!25}6.05$\times10^{-7}$& \cellcolor{green!25} 2.28$\times10^{-6}$ \\
GEE (Avg) test set  &  \cellcolor{green!25}1.08$\times10^{-6}$& \cellcolor{green!25} 1.81$\times10^{-6}$ & \cellcolor{green!25}6.05$\times10^{-7}$& \cellcolor{green!25} 2.25$\times10^{-6}$ \\
\hline
\end{tabular}
\caption{MSE comparison between the GEE and LL models on predicted ASDRs ($q_{cgxt}$) for AUT and CZE.}
\label{tab3}
\end{table}

The proportion of variance explained by both $\rm{PC1}^{[20,50]}$ and $\rm{PC1}^{[51,80]}$ is 99 percent. The high proportion of variance explained by PC1 indicates that it is a powerful summary measure of the mortality data, and incorporating it into the GEE model has led to improved forecasting performance. It suggests that the mortality rates for age groups 20 to 80 exhibit strong patterns or trends that are effectively captured by PC1, enabling more accurate predictions of future mortality rates.

\section{Modelling Mortality Rates in Single Population}\label{sec3}

We introduce two approaches for defining a covariate $k_{t}$ in the context of modeling and forecasting mortality rates within single-population settings. The first approach, PCA-GEE, employs principal component analysis to derive this covariate, while the second approach, Avg-GEE, entails averaging mortality rates across predefined age groups (as described in \cite{dastranj2023age}). Our investigation will evaluate both PCA-GEE and Avg-GEE methods and compare their forecast accuracy against that of the LC model. This analysis aims to identify the most effective approach for modeling and forecasting mortality rates.

We now begin to establish the PCA-GEE method for a single population. Let $m_{x,t}$ denote the ASDR at age $x$ and time $t$, for $x = 0, 1, \cdots , \omega$; and $t=0,1, \cdots, T$ \citep{dickson2019actuarial, macdonald2018modelling}. Let $y_{xt}= \log (m_{x,t})$. 

We categorize individuals into three distinct age groups: 0 to 19, 20 to 50, and 51 to $\omega$, treating each group as homogeneous. These age categories are identified as children, young adults, and older adults, respectively. 

We conduct PCA on the children, young adults, and older adults within the dataset. Subsequently, we define covariate $k_{t}$ as follows:

\begin{align}
k_{t} = &
\begin{cases} 
{\rm PC1}_{t}^{[0,19]} & \text{if}\ x  \in [0,19], \\ 
{\rm PC1}_{t}^{[20,50]} & \text{if}\ x  \in [20,50], \\ 
      {\rm PC1}_{t}^{[51,\omega]} & {\rm otherwise},
\end{cases}
\label{ali5}
\end{align}
for $t=0, 1,  \cdots, T$. 

We will employ random walks with drift to forecast the future values of $k_{t}$ across different age ranges: ages $0$ to $19$, ages $20$ to $50$, and ages $51$ to $\omega$.

The PCA-GEE model for single-population scenarios, executed via the geeglm function within the R geepack package (Højsgaard et al., 2006), is specified as follows:

\begin{align}
    & \text{geeglm}(y \sim \text{age} + \text{age:I}(k_{t}) + \text{age:I}(k_{t}^2) + \text{cohort}, \nonumber \\
    & \qquad \text{id} = \text{age}, \text{waves} = \text{year}, \text{corstr} = \text{``ar1''}, \nonumber \\
    & \qquad \text{weights} = \text{agenum}/\text{mean(agenum)}, \text{data} = \text{ASDRs}) \nonumber
\end{align}

Mathematically, the PCA-GEE model for the single-population scenario can be expressed as:

\begin{equation}\label{eq3}
    y_{xt} = a_x + b_xk_{t} + c_xk_{t}^2 + \gamma(t-x) + \epsilon_{xt}
\end{equation}
Here, $a_x$ denotes the age effect of the intercept, while $b_x$ and $c_x$ represent the age effects of $k_{t}$ and $k_{t}^2$, respectively. The term $\gamma$ denotes the cohort effect, and $\epsilon_{xt}$ represents the error term. The term $\gamma$ demonstrates the influence of birth cohorts on mortality rates. The cohort effect suggests that individuals born in distinct years (cohorts) may experience varying mortality rates due to unique historical, societal, or environmental factors relevant to their birth years.

We will now delve into the methodology of Avg-GEE modelling. Let $m_{x,t}$ denote the ASDR at age $x$ and time $t$, for $x = 0,1, \cdots , \omega$ and $t=0,1, \cdots, T$, see \citet{dickson2019actuarial}. Let $y_{xt}= \log (m_{x,t})$, the log of the observed age-specific death rates in a given year $t$. Let $k_{t}$ represent the average of $y_{xt}$ for two distinct age range groups. These groups include the average for age groups 0 to 19, the average for age groups 20 to 50, and the average for age groups 51 to $\omega$:

\begin{align}
k_{t} = &
\begin{cases} 
k_{t}^{[0,19]} & \text{if}\ x  \in [0,19], \\ 
k_{t}^{[20,50]} & \text{if}\ x  \in [20,50], \\ 
      k_{t}^{[51,\omega]} & {\rm otherwise},
\end{cases}
\label{ali7}
\end{align}
for $t=0, 1,  \cdots, T$, where
\begin{align*}
k_{t}^{[0,19]}=
\dfrac{\sum\limits_{x=0}^{19} y_{xt}}{ 20}  , \qquad k_{t}^{[20,50]}=
\dfrac{\sum\limits_{x=20}^{50} y_{xt}}{31}  , \qquad
k_{t}^{[51,\omega]}=\dfrac{\sum\limits_{x=51}^{\omega} y_{xt}}{\omega-50},
\end{align*}
 for $t=0, 1,  \cdots, T$. $k_{t}$ will be included as predictor in our Avg-GEE model. We will utilize random walks with drift to predict the future values of $k_{t}$ for ages $0$ to $19$, as well as the future values of $k_{t}$ for ages $20$ to $50$ and ages $51$ to $\omega$.

 The Avg-GEE model for single-population scenarios, implemented through the geeglm function in the geepack package for R (Højsgaard et al., 2006), follows a similar definition to the PCA-GEE model for single populations. However, in the Avg-GEE model, we utilize the covariate $k_t$, as defined in Equation \ref{ali5}, which is derived by averaging the logarithm of mortality rates across predefined age groups.

\subsection{Single Population Analysis Across 56 Populations: Ages 20-80}

In this section, our analysis extends to encompass a broader dataset comprising fifty-six populations. This data, sourced from the HMD, is organized into one-year age intervals (1x1 intervals). Our objective is to individually examine the ASDRs of these fifty-six populations and discern any discernible patterns or trends.

For our analysis, we focus on age groups spanning from 20 to 80 years. The mortality rates from 1991 to 2010 serve as our training dataset, enabling us to estimate the parameters of the ASDRs models. Subsequently, for the period spanning from 2011 to 2019 (prior to the onset of the Covid-19 pandemic), we forecast the ASDRs for each individual population.

Our approach entails applying a specific variant of the GEE model, denoted as model \eqref{eq3}, to each unique population. This variant does not account for cohort effects and is expressed as follows:

\begin{equation}\label{eq:4}
    y_{xt} = a_x + b_xk_{t} + \epsilon_{xt}
\end{equation}

The equation \eqref{eq:4} and the LC model may appear similar in structure, but their conceptual frameworks and interpretations are distinct. In equation \eqref{eq:4}, \( k_t \) represents the known mortality covariate derived from equation \eqref{ali5}. The parameters \( a_x \) and \( b_x \) signify the age effect on the intercept and slope, respectively.

Our analysis involve dividing the MSEs of LC by PCA-GEE and Avg-GEE MSEs. A ratio greater than one signifies superior forecast accuracy of the GEE model for a given population. Our findings indicate that for 29 out of 56 populations, the PCA-GEE models exhibit lower MSEs compared to LC models, while for 30 populations, the Avg-GEE models outperform the LC models. This highlights the superior performance of GEE models over the LC model, showcasing their efficiency in accurately predicting mortality rates across diverse populations.

\begin{figure}[H]
    \centering
    \includegraphics[width=\textwidth]{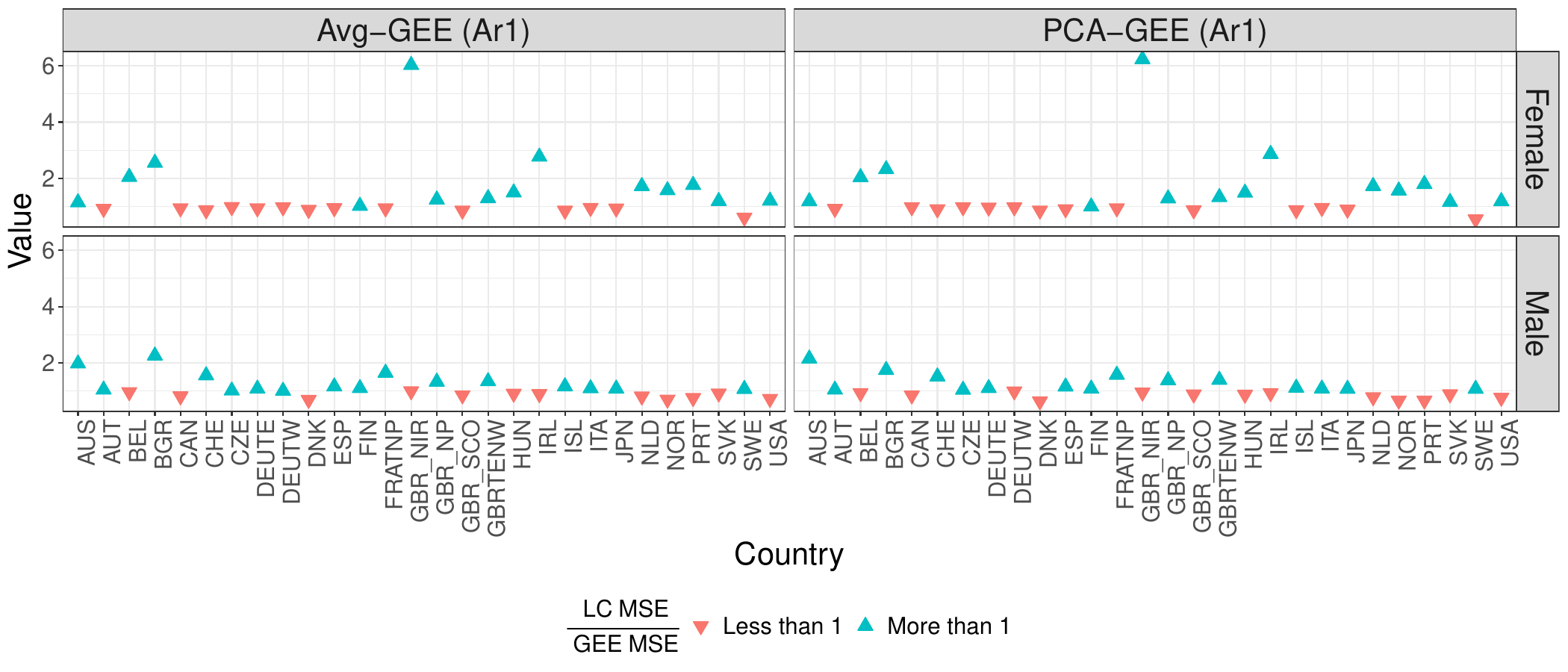}
    \caption{Comparison of MSEs between the GEE and LC models in predicting ASDRs ($m_{xt}$) across 56 populations.}
    \label{fig2}
\end{figure}

The R code for fitting and forecasting the GEE models to mortality data is available in a GitHub repository maintained by the first author of this paper \citep{DastranjR}. Specifically, the code covers the case of multi-populations for both males and females in Austria and the Czech Republic.

\section{Conclusion}

This study investigated the application of GEEs in conjunction with PCA for analyzing ASDRs in longitudinal datasets. GEEs demonstrated considerable promise in this context due to their ability to handle correlated data, a common feature in mortality studies.  The computational efficiency of GEEs makes them a valuable tool, particularly when dealing with categorical data. 

Our findings suggest that GEEs combined with PCA offer a powerful approach for analyzing longitudinal ASDR data. This approach can contribute to a deeper understanding of mortality trends and enhance the quality of research in this crucial field. By leveraging the strengths of GEEs and PCA, we can gain valuable insights into mortality patterns and ultimately improve public health initiatives focused on longevity.

\bibliographystyle{chicago}
\bibliography{mybib}

\begin{thebibliography}{}

\bibitem[\protect\citeauthoryear{Abdi and Williams}{Abdi and Williams}{2010}]{abdi2010principal}
Abdi, H. and L.~J. Williams (2010).
\newblock Principal component analysis.
\newblock {\em Wiley interdisciplinary reviews: computational statistics\/}~{\em 2\/}(4), 433--459.

\bibitem[\protect\citeauthoryear{Booth and Tickle}{Booth and Tickle}{2008}]{booth2008mortality}
Booth, H. and L.~Tickle (2008).
\newblock Mortality modelling and forecasting: A review of methods.
\newblock {\em Annals of actuarial science\/}~{\em 3\/}(1-2), 3--43.

\bibitem[\protect\citeauthoryear{Box, Jenkins, Reinsel, and Ljung}{Box et~al.}{2015}]{box2015time}
Box, G.~E., G.~M. Jenkins, G.~C. Reinsel, and G.~M. Ljung (2015).
\newblock {\em Time series analysis: forecasting and control}.
\newblock John Wiley \& Sons.

\bibitem[\protect\citeauthoryear{Burger, Baudisch, and Vaupel}{Burger et~al.}{2012}]{burger2012human}
Burger, O., A.~Baudisch, and J.~W. Vaupel (2012).
\newblock Human mortality improvement in evolutionary context.
\newblock {\em Proceedings of the National Academy of Sciences\/}~{\em 109\/}(44), 18210--18214.

\bibitem[\protect\citeauthoryear{Dastranj and Kolář}{Dastranj and Kolář}{2023}]{dastranj2023age}
Dastranj, R. and M.~Kolář (2023).
\newblock Age-gender-country-specific death rates modelling and forecasting: A linear mixed-effects model.
\newblock {\em arXiv preprint arXiv:2311.18668\/}.

\bibitem[\protect\citeauthoryear{Dastranj and Kolář}{Dastranj and Kolář}{2024}]{DastranjR}
Dastranj, R. and M.~Kolář (2024).
\newblock {PCAGEE}.
\newblock \url{https://rezadastranj.github.io/PCA-GEE/}.

\bibitem[\protect\citeauthoryear{Dickson, Hardy, and Waters}{Dickson et~al.}{2019}]{dickson2019actuarial}
Dickson, D.~C., M.~R. Hardy, and H.~R. Waters (2019).
\newblock {\em Actuarial mathematics for life contingent risks}.
\newblock Cambridge University Press.

\bibitem[\protect\citeauthoryear{Frees}{Frees}{2004}]{frees2004longitudinal}
Frees, E.~W. (2004).
\newblock {\em Longitudinal and panel data: analysis and applications in the social sciences}.
\newblock Cambridge University Press.

\bibitem[\protect\citeauthoryear{Gelman}{Gelman}{2005}]{gelman2005analysis}
Gelman, A. (2005).
\newblock Analysis of variance—why it is more important than ever.
\newblock {\em The Annals of Statistics\/}~{\em 33\/}(1), 1--53.

\bibitem[\protect\citeauthoryear{Gompertz}{Gompertz}{1825}]{gompertz1825xxiv}
Gompertz, B. (1825).
\newblock Xxiv. on the nature of the function expressive of the law of human mortality, and on a new mode of determining the value of life contingencies. in a letter to francis baily, esq. frs \&c.
\newblock {\em Philosophical transactions of the Royal Society of London\/}~(115), 513--583.

\bibitem[\protect\citeauthoryear{Green~Jr and Tukey}{Green~Jr and Tukey}{1960}]{green1960complex}
Green~Jr, B.~F. and J.~W. Tukey (1960).
\newblock Complex analyses of variance: general problems.
\newblock {\em Psychometrika\/}~{\em 25\/}(2), 127--152.

\bibitem[\protect\citeauthoryear{Hardin and Hilbe}{Hardin and Hilbe}{2002}]{hardin2002generalized}
Hardin, J.~W. and J.~M. Hilbe (2002).
\newblock {\em Generalized estimating equations}.
\newblock chapman and hall/CRC.

\bibitem[\protect\citeauthoryear{HMD}{HMD}{2022}]{HMD}
HMD (2022).
\newblock {Human Mortality Database. Max Planck Institute for Demographic Research (Germany), University of California, Berkeley (USA), and French Institute for Demographic Studies (France)}.
\newblock Available at \url{www.mortality.org} or \url{www.humanmortality.de} (data downloaded on 09-05-2022].

\bibitem[\protect\citeauthoryear{H{\o}jsgaard, Halekoh, and Yan}{H{\o}jsgaard et~al.}{2006}]{hojsgaard2006r}
H{\o}jsgaard, S., U.~Halekoh, and J.~Yan (2006).
\newblock The r package geepack for generalized estimating equations.
\newblock {\em Journal of statistical software\/}~{\em 15}, 1--11.

\bibitem[\protect\citeauthoryear{James, Witten, Hastie, Tibshirani, et~al.}{James et~al.}{2013}]{james2013introduction}
James, G., D.~Witten, T.~Hastie, R.~Tibshirani, et~al. (2013).
\newblock {\em An introduction to statistical learning}, Volume 112.
\newblock Springer.

\bibitem[\protect\citeauthoryear{Jolliffe}{Jolliffe}{2002}]{jolliffe2002principal}
Jolliffe, I.~T. (2002).
\newblock {\em Principal component analysis}.
\newblock New York; Springer.

\bibitem[\protect\citeauthoryear{Kleinow}{Kleinow}{2015}]{kleinow2015common}
Kleinow, T. (2015).
\newblock A common age effect model for the mortality of multiple populations.
\newblock {\em Insurance: Mathematics and Economics\/}~{\em 63}, 147--152.

\bibitem[\protect\citeauthoryear{Lee and Carter}{Lee and Carter}{1992}]{lee1992modeling}
Lee, R.~D. and L.~R. Carter (1992).
\newblock Modeling and forecasting {U.S.} mortality.
\newblock {\em Journal of the American statistical association\/}~{\em 87\/}(419), 659--671.

\bibitem[\protect\citeauthoryear{Lee and Nelder}{Lee and Nelder}{2004}]{lee2004conditional}
Lee, Y. and J.~A. Nelder (2004).
\newblock Conditional and marginal models: another view.

\bibitem[\protect\citeauthoryear{Li and Lee}{Li and Lee}{2005}]{li2005coherent}
Li, N. and R.~Lee (2005).
\newblock Coherent mortality forecasts for a group of populations: An extension of the {Lee-Carter} method.
\newblock {\em Demography\/}~{\em 42\/}(3), 575--594.

\bibitem[\protect\citeauthoryear{Liang and Zeger}{Liang and Zeger}{1986}]{liang1986longitudinal}
Liang, K.-Y. and S.~L. Zeger (1986).
\newblock Longitudinal data analysis using generalized linear models.
\newblock {\em Biometrika\/}~{\em 73\/}(1), 13--22.

\bibitem[\protect\citeauthoryear{Macdonald, Richards, and Currie}{Macdonald et~al.}{2018}]{macdonald2018modelling}
Macdonald, A.~S., S.~J. Richards, and I.~D. Currie (2018).
\newblock {\em Modelling mortality with actuarial applications}.
\newblock Cambridge University Press.

\bibitem[\protect\citeauthoryear{McCullagh and Nelder}{McCullagh and Nelder}{1989}]{mccullagh2019generalized}
McCullagh, P. and J.~Nelder (1989).
\newblock {\em Generalized Linear Models. 2nd Edition}.
\newblock Chapman and Hall, London.

\bibitem[\protect\citeauthoryear{Nelder and Wedderburn}{Nelder and Wedderburn}{1972}]{nelder1972generalized}
Nelder, J.~A. and R.~W. Wedderburn (1972).
\newblock Generalized linear models.
\newblock {\em Journal of the Royal Statistical Society Series A: Statistics in Society\/}~{\em 135\/}(3), 370--384.

\bibitem[\protect\citeauthoryear{Pan and Connett}{Pan and Connett}{2002}]{pan2002selecting}
Pan, W. and J.~E. Connett (2002).
\newblock Selecting the working correlation structure in generalized estimating equations with application to the lung health study.
\newblock {\em Statistica Sinica\/}, 475--490.

\bibitem[\protect\citeauthoryear{Pinheiro and Bates}{Pinheiro and Bates}{2006}]{pinheiro2006mixed}
Pinheiro, J. and D.~Bates (2006).
\newblock {\em Mixed-Effects models in S and S-PLUS}.
\newblock Springer Science \& Business Media.

\bibitem[\protect\citeauthoryear{Richman and W{\"u}thrich}{Richman and W{\"u}thrich}{2021}]{richman2021neural}
Richman, R. and M.~V. W{\"u}thrich (2021).
\newblock A neural network extension of the {Lee-Carter} model to multiple populations.
\newblock {\em Annals of Actuarial Science\/}~{\em 15\/}(2), 346--366.

\end{thebibliography}

\end{document}